\documentclass[%
 reprint,
 amsmath,
 amssymb,
 aps,
 floatfix
]{revtex4-2}
\usepackage[english]{babel}
\usepackage{graphicx}
\usepackage{dcolumn}
\usepackage{bm}
\usepackage{siunitx}
\usepackage{comment}
\usepackage{svg}
\usepackage[hidelinks]{hyperref}
\usepackage{float}
\usepackage{braket}
\DeclareSIUnit\barn{b}
\makeatletter
\let\ORIbbl@fixname\bbl@fixname
\def\bbl@fixname#1{%
  \@ifundefined{languagealias@\expandafter\string#1}
    {\ORIbbl@fixname#1}
    {\edef\languagename{\@nameuse{languagealias@#1}}}%
}
\newcommand{\definelanguagealias}[2]{%
  \@namedef{languagealias@#1}{#2}%
}
\makeatother
\definelanguagealias{en}{english}
\definelanguagealias{En}{english}
\definelanguagealias{EN}{english}
\begin{document}
\preprint{APS/123-QED}
\title{Indirect measurement of the carbon adatom migration barrier on graphene}
\author{Andreas Postl}
\email{andreas.postl@univie.ac.at}
\author{Pit Pascal Patrick Hilgert}
\author{Alexander Markevich}
\author{Jacob Madsen}
\author{Kimmo Mustonen}
\author{Jani Kotakoski}
\author{Toma Susi}
\email{toma.susi@univie.ac.at}
\affiliation{University of Vienna, Faculty of Physics, Boltzmanngasse 5, A-1090 Vienna}
\date{\today}
\begin{abstract}
Although surface diffusion is critical for many physical and chemical processes, including the epitaxial growth of crystals and heterogeneous catalysis, it is particularly challenging to directly study. Here, we estimate the carbon adatom migration barrier on freestanding monolayer graphene by quantifying its temperature-dependent electron knock-on damage. Due to the fast healing of vacancies by diffusing adatoms, the damage rate decreases with increasing temperature. By analyzing the observed damage rates at 300--1073~\si{\kelvin} using a model describing our finite scanning probe, we find a barrier of $(0.33 \pm 0.03)~\si{\electronvolt}$.
\end{abstract}
%
%
\maketitle
Transmission electron microscopy (TEM) allows exposing specimens to electrons impinging with high kinetic energy (typically up to 200--300~\si{\kilo\electronvolt}) and imaging the effects in-situ with atomic resolution. Recent work has established that electron irradiation can be used to sculpt materials~\cite{lin_flexible_2014, ryu_atomic-scale_2015, wang_atomic_2017, gilbert_fabrication_2017, clark_scalable_2018}, induce phase transitions~\cite{lin_vacancy-induced_2015, liu_phase_2017}, locally amorphize~\cite{eder_journey_2014, su_-situ_2018} or crystallize~\cite{jesse_atomic-level_2015, bayer_atomic-scale_2018} structures, and even to manipulate individual covalently bound atoms~\cite{susi_siliconcarbon_2014, susi_manipulating_2017, dyck_placing_2017, hudak_directed_2018}. Diffusion processes of fundamental importance have also been directly studied in some materials, although unavoidably these observations have been influenced by the energetic electron beam~\cite{ishikawa_direct_2014,li_column-by-column_2016,furnival_anomalous_2017}. 

Understanding the interaction between probe electrons and the sample has become crucial to correctly apply and interpret such experiments~\cite{susi_quantifying_2019}. The investigation of irradiation effects in carbon nanostructures has been a field of intense research during the last decades~\cite{banhart_irradiation_1999,zobelli_electron_2007,banhart_structural_2011}. Recently, progress in sample preparation of two-dimensional materials and advances in the theoretical models have, especially in graphene, enabled the quantitative description of so-called knock-on damage resulting from elastic electron-nucleus collisions enhanced by atomic vibrations~\cite{meyer_accurate_2012, susi_isotope_2016}, whereas inelastic scattering and its contribution to damage are still harder to describe~\cite{susi_quantifying_2019, kretschmer_formation_2020}.
However, with notable exceptions (albeit not at atomic resolution~\cite{cretu_inelastic_2015}), thus far the effect of temperature on such processes has been rarely quantified.

In this study, our initial aim was to determine the temperature-dependence of the electron knock-on damage cross section for pristine graphene, which is in the range of 5--20~\si{\milli\barn} for $90$~\si{\kilo\electronvolt} electrons and $^{12}$C lattice atoms at ambient temperature~\cite{susi_isotope_2016}. Based on a first-principles model of the cross section, one should expect to observe tremendously increasing knock-on damage rates for elevated temperatures due to the higher population of out-of-plane phonon modes~\cite{susi_isotope_2016} and the thermal perturbation of the lattice~\cite{chirita_mihaila_influence_2019}. In stark contrast to that prediction, the detected damage rates do not increase with temperature, but rather decrease.

The reason must be thermally activated carbon adatom migration and recombination with defects. This has been directly observed for vacancies and larger holes in graphene~\cite{zan_graphene_2012,susi_isotope_2016} as well as at its impurity sites~\cite{tripathi_implanting_2018,su_engineering_2019}, even at room temperature, and also indirectly studied for carbon nanotubes under electron irradiation~\cite{gan_diffusion_2008}. A carbon adatom on top of a graphene layer bonds at a C--C bridge site and has to overcome an energy barrier, estimated to be in the range of 0.40--0.47~\si{\electronvolt}~\cite{lehtinen_magnetic_2003, krasheninnikov_adsorption_2004}, to migrate from one minimum to another. At elevated temperatures, this migration is enhanced, and thus the proportion of vacancies which get healed before they are detected increases with increasing temperature.

Here, we are able to use the discrepancy between predicted and observed damage rates of electron knock-on damage by $90$~\si{\kilo\electronvolt} electrons at elevated temperatures between $300$ and $1073$~\si{\kelvin} to provide an indirect experimental estimate of the migration barrier. Importantly, we need to account for the fact that the same scanning electron beam both creates and observes the damage to correctly describe the experiments. Our analysis indicates a barrier value of $(0.33 \pm 0.03)~\si{\electronvolt}$ (in line with the $0.25~\si{\electronvolt}$ estimated inside nanotubes~\cite{gan_diffusion_2008}), which is the first measurement for graphene that has been reported to date.

As samples, we used commercial monolayer graphene (Easy Transfer, Graphenea S.A.), which was transferred onto a chip with an electron-transparent window and electrical contacts for resistive heating, and placed in an in-situ TEM holder with an integrated electrical circuit (Fusion, Protochips Inc.). To heat the sample, a current was passed through the heating coil of the chip. The heating power and temperature were controlled based on the manufacturer's per-chip calibration, and the precision of the set temperature was estimated to be $\pm$2~\%.

All experimental images were acquired using a Nion \mbox{UltraSTEM} 100, a probe-corrected dedicated scanning transmission electron microscope (STEM)~\cite{krivanek_monochromated_2013}, operated at 90 keV with a probe convergence semi-angle of 30 mrad. Crucially, our column pressure is near ultra-high vacuum ($\lesssim 1\times10^{-9}$~mbar), which minimizes any spurious effects of chemical etching. We scanned across fields of view of roughly $1 \times 1$~\si{\nano\meter}$^2$ or $2 \times 2$~\si{\nano\meter}$^2$, which initially contained pristine graphene and were located away from any surface contamination, and recorded medium-angle annular dark field (MAADF; 60--200~\si{\milli\radian} collection semi-angle) image series of consecutive frames.

\begin{figure}[t]
  \centering
  \includegraphics[width=\columnwidth]{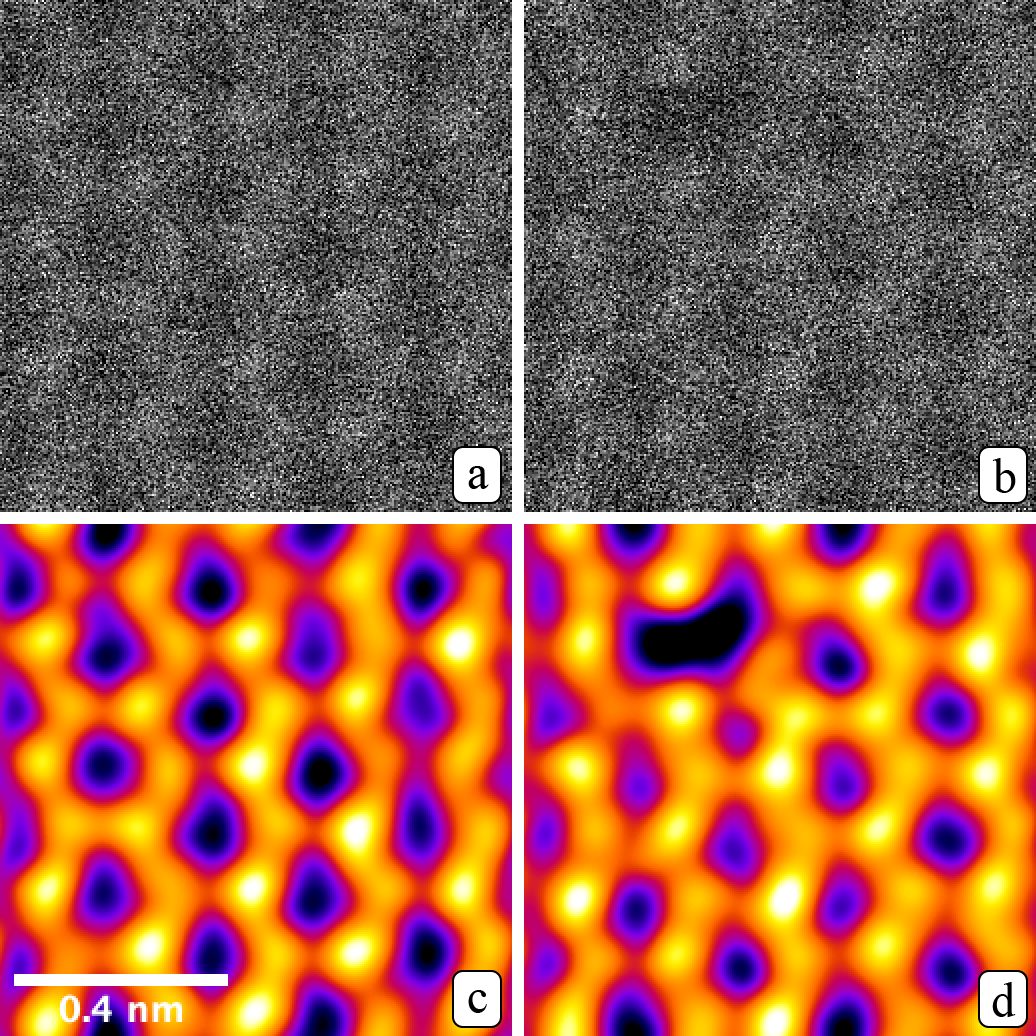}
  \caption{Raw (a, b) and colored double Gaussian-filtered (c, d) medium-angle ADF-STEM images before (a, c) and after (b, d) a knock-on damage event at $100$~\si{\celsius}. Field of view: $1 \times 1$~\si{\nano\meter}$^2$, pixel dwell time: $16$~\si{\micro\second}, $256 \times 256$ px}
  \label{fig:ADFSTEM}
\end{figure}

As illustrated in Fig.~\ref{fig:ADFSTEM}, we stopped the acquisition whenever we recognized a defect that did not conserve the number of atoms (as opposed to e.g. a Stone-Wales 5577 defect~\cite{stone_theoretical_1986, kotakoski_stone-wales-type_2011}). We scanned as quickly as we could while retaining atomic resolution, and to enhance contrast, used double-Gaussian filtering~\cite{krivanek_gentle_2010} of the raw images during acquisition. However, if we found (either during the acquisition or the later analysis) that such a SW 5577 defect was immediately followed by a defect that did not conserve the number of atoms, we excluded that series from any further evaluations, as the local threshold energy for the vacancy creation would not correspond to that of the pristine lattice.

To accurately estimate the beam current during imaging, we related it to the current from electrons hitting the virtual objective aperture (VOA) of our STEM, which is recorded when images are taken (for the calibration curve, see \cite{supplement}). The beam current as a function of the VOA current was recorded at least every other week when the experiments were conducted.

At ambient temperature, the recombination rate of carbon adatoms and vacancies is low compared to the rate of further atom loss under irradiation with typical beam currents of 50--100~\si{\pico\ampere} at $90$~\si{\kilo\electronvolt}. At elevated temperatures, the recombination rate rapidly rises and greatly exceeds even our highest frame acquisition rate at an acceptable signal-to-noise ratio limit, which was about two frames per second (pixel dwell time: $8$~\si{\micro\second}, $256 \times 256$ pixels) in our setup (for estimated rates, see \cite{supplement}). Thus, graphene samples mainly show increased radiation hardness at elevated temperatures, as has been remarked before~\cite{song_atomic-scale_2011, zan_graphene_2012}. However, this effect was not quantified until now.

After obtaining sufficient data to perform statistical analyses ($60$ or more series per temperature and frame acquisition time), we examined each recorded image series to identify the first frame in which at least one atom was missing (henceforth referred to as defect frame). The total electron count $N_{\mathrm{e}^-}$ up to that point was calculated with the assistance of image metadata and the above-mentioned beam current calibration. For the defect frame itself, we counted half of the frame time and neglected the $(x, y)$ position of the defect in the frame. These data were analyzed as a homogeneous Poisson process~\cite{kingman1992poisson} (for detail, see \cite{supplement}). We did not observe a difference with respect to a variation of frame acquisition rate from $2$ to $0.5$ fps. Thus, we merged our data for each temperature.

Although it turned out that our measurement cannot be used to calculate the true temperature-dependent knock-on cross section, in the following we use the term "observed cross section" with the symbol $\sigma^\mathrm{obs}_\mathrm{ko}$. 
To begin with, we slightly revised the parameters of our knock-on damage cross section model, incorporating our additional $90$~\si{\kilo\electronvolt} room-temperature data. Due to possible phonon modeling inaccuracies, we used a parameter uncertainty for the out-of-plane root-mean-square velocity of the nuclei $v_\mathrm{rms}(T)$ and refitted the threshold energy $T_\mathrm{d}$ (for detail, see \cite{supplement}). Variance-weighted least squares with a trust region reflective algorithm~\cite{conn2000trust} yielded $T_\mathrm{d} = (21.03 \pm 0.10)~\si{\electronvolt}$ with $v_\mathrm{rms}(300~\mathrm{K}) = (590 \pm 20)~\si{\meter\per\second}$.
\begin{figure*}[t]
\includegraphics[width=0.86\textwidth, keepaspectratio]{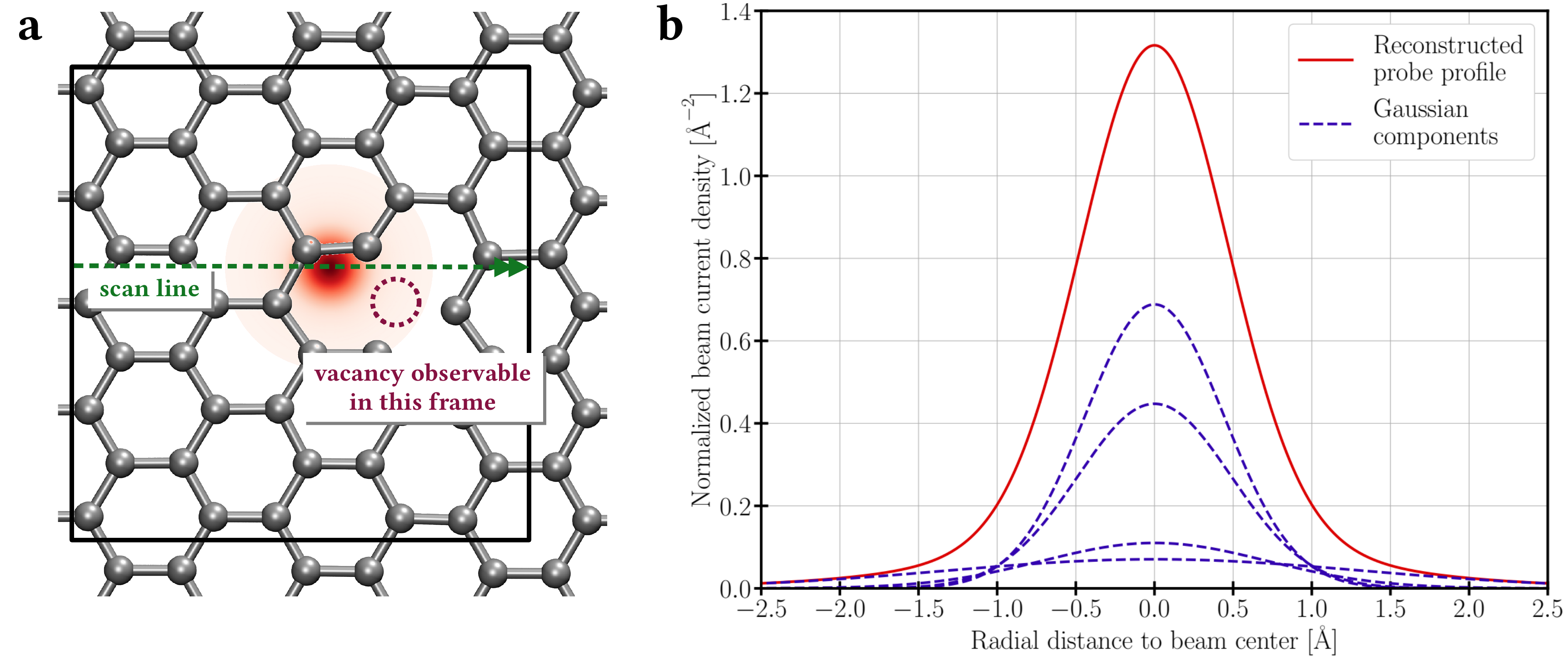}
\caption{(a) Schematic illustration of the graphene lattice and a single vacancy (dashed purple circle) created by the "leading" tail of the scanning electron probe and observed within the same scan frame (black rectangle). (b) Radial plot of the probe current density profile fitted by four Gaussian components.}
\label{fig:schematics_probe}
\end{figure*}

By assuming that the discrepancy between the experimentally observed knock-on damage rate $k_\mathrm{ko}^\mathrm{obs}(I_\mathrm{B}, T)$ and the predicted rate $k_\mathrm{ko}^\mathrm{{theor}} = (I_\mathrm{B}/e) \, \rho_A \, \sigma^\mathrm{theor}_\mathrm{ko}$, with $I_B$ being the beam current, $e$ the elementary charge and $\rho_A$ the areal atomic density of graphene, is equal to the healing rate of vacancies $k_\mathrm{h}(T)$,
\begin{equation}
    k_\mathrm{ko}^\mathrm{{theor}}\left( I_\mathrm{B}, T \right)
    -
    k_\mathrm{ko}^\mathrm{{obs}}(T)
    =
    k_\mathrm{h}(T),
\label{eq:kh1}
\end{equation}
we can state an Arrhenius dependence~\cite{arrhenius_uber_1889} of the healing rate on the migration energy barrier $E_\mathrm{m}$ as
\begin{equation}
    k_\mathrm{h}(T)
    \overset{E_\mathrm{r}=0}{\approx}
    A
    \exp \left( 
        - \frac{E_\mathrm{m}}{k_\mathrm{B} T}
    \right),
\label{eq:kh2}
\end{equation}
where $A$ is the pre-exponential rate constant, $T$ the absolute temperature, $k_\mathrm{B}$ the Boltzmann constant, and $E_\mathrm{r}$ the energy barrier for the recombination of a vacancy and a carbon adatom in immediate proximity, which we assume to be negligible compared to $E_\mathrm{m}$. This naive treatment yields $E_\textrm{m} = (150 \pm 6)~\si{\milli\electronvolt}$ for the adatom migration barrier (see \cite{supplement}). The resulting pre-exponential factor is close to the frame acquisition frequency, which underscores a limitation of the applied measuring method; since we counted only one knock-on event per vacancy irrespective of its size, the maximum observed damage rate would be equal to the frame rate. Eq.~\eqref{eq:kh2} would hold if $k_\mathrm{ko}^\mathrm{{theor}}$ were much higher than the healing rate $k_\mathrm{h}(T)$. For temperatures above $400~\si{\kelvin}$, however, the observed cross section values are very low ($\lesssim 5$ mb). 
Despite the fact that the healing of vacancies is on average much faster than knock-on damage for temperatures up to $1073~\si{\kelvin}$, we are occasionally able to observe them, namely if a created vacancy is not healed before it can be observed (for modeled rates, see \cite{supplement}).

To correctly describe these observations, we must explicitly account for the nature of the experiment: the images are recorded by a scanning electron probe with a finite current density distribution (Fig.~\ref{fig:schematics_probe}). Thus, the time between the creation of a vacancy and its observation depends on where it is created with respect to the probe. This motivates an extension of the reduced healing rate model of Eq.~\eqref{eq:kh2} to explicitly account for this probability.

To start with, we redefine the healing rate as a fraction of the theoretical knock-on rate determined by probability $P_\textrm{h} (T)$ for a vacancy to be healed before observation
\begin{equation}
    k_\textrm{h} (T) = P_\textrm{h} (T) k_\textrm{ko}^\textrm{theor} (T),
    \label{eq:kh3}
\end{equation}
which can be combined with Eq.~\eqref{eq:kh1} to obtain a new effective observed damage rate
\begin{equation}
    k_\textrm{ko}^\textrm{obs} (T) = k_\textrm{ko}^\textrm{theor} (T) \left(
    1 - P_\textrm{h} (T)
    \right).
    \label{eq:kh4}
\end{equation}
An accurate description of $P_\textrm{h} (T)$ must contain the involved random variables via their probability distributions. A defect can not be observed if the number of adatom migration steps within a frame time ($n_\textrm{f}(T)$) is greater than the number of steps needed to reach a vacancy to heal it ($n_\textrm{h}$). Specifically, the healing probability is the value of the complementary cumulative distribution function (tail distribution) of the random variable $Q(T) = n_\textrm{f}(T) / n_\textrm{h}$ at 1. 
The number of migration steps is normally distributed with a mean of
\begin{equation}
    \mu(T) = t_\textrm{f} f_0 \exp \left( - \frac{E_\textrm{m}}{k_\textrm{B} T} \right),
    \label{eq:normdist-mu}
\end{equation}
where $t_\textrm{f}$ is the frame time, and $f_0$ the migration attempt frequency ($4 \times 10^{12}~\mathrm{s}^{-1}$ as reported for carbon interstitials in graphite \cite{thrower_point_1978}).
The number of surface diffusion steps that $N_\textrm{v}$ adatoms need to reach the immediate proximity of a vacancy with (on average) $N_\textrm{v}$ missing atoms is exponentially distributed with the parameter
\begin{equation}
    \nu = - \frac{1}{N_\textrm{v}} \log \left( 1-\frac{2}{3} c_\textrm{ad} \right),
    \label{eq:nu}
\end{equation}
where $c_\textrm{ad}$ is the number of adatoms per lattice atom (adatom concentration), and the prefactor $2/3$ accounts for the number of bonds per atom. In our data, the average vacancy size is $N_\textrm{v} = 1.6\pm0.2$ (see \cite{supplement}).

\begin{figure}[t!]
  \centering
  \includegraphics[width=\columnwidth]{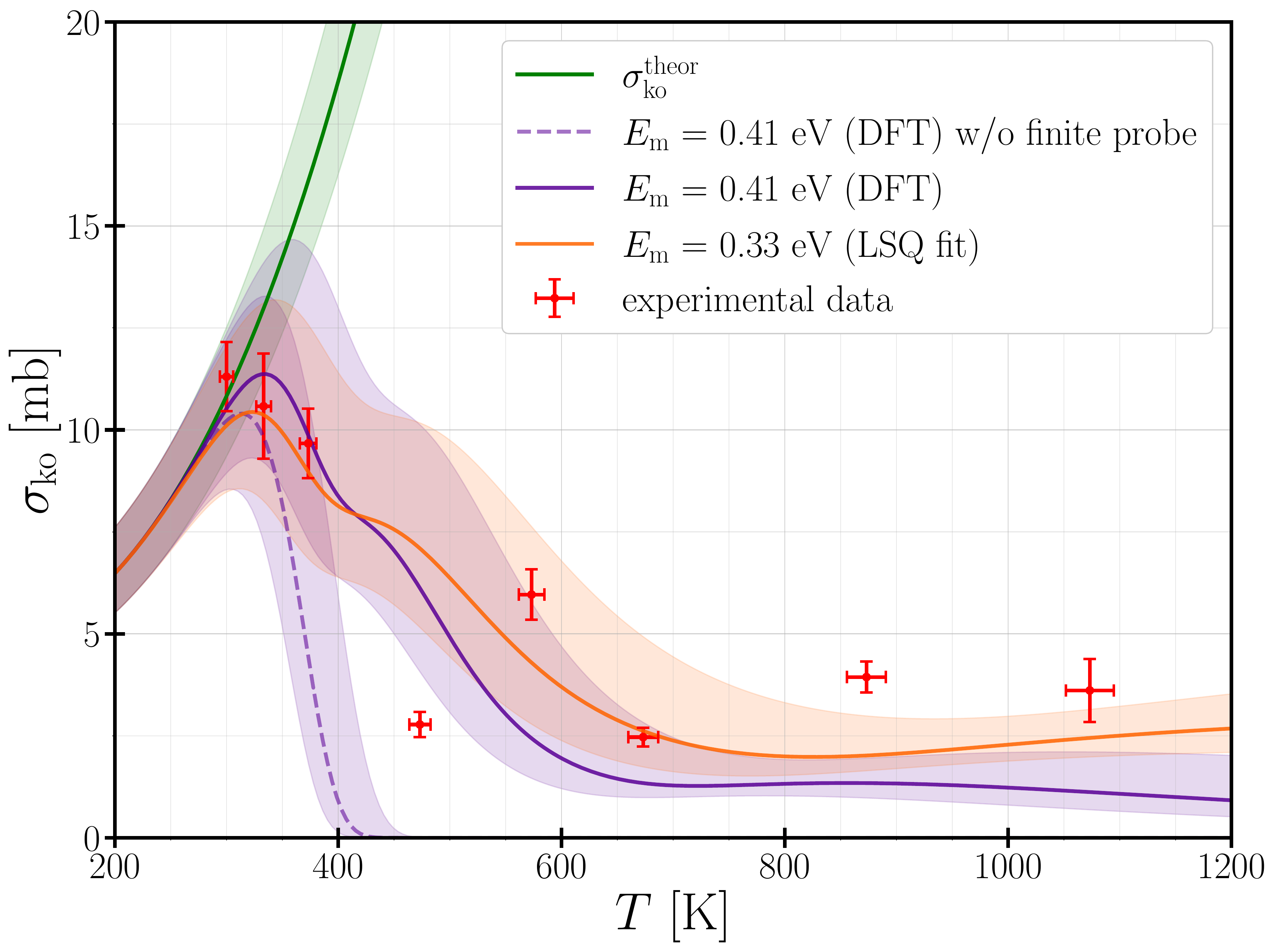}
  \caption{Knock-on damage cross section as a function of temperature: experimental observation (red points), theoretical model (green line), and an extended model describing the damage rate with counteracting vacancy healing and the effect of a scanning electron probe with a finite extent, resulting in a varying detection time, and the measured shape (orange line for a least-squares fitted migration barrier and purple for the DFT barrier). See the main text for description of the underlying adatom concentrations and error estimates.}
  \label{fig:cross_section_full_model}
\end{figure}

The tail distribution of the ratio $Q(T)$ can be approximated (see \cite{supplement}) by the cumulative distribution function of $n_\textrm{h}$ at the expectation value of $n_\textrm{f}(T)$,
\begin{equation}
    P_\textrm{h} (T) 
    = \int_0^{\mu (T)} \nu \exp \left( - \nu k \right) \mathrm{d}k 
    = 1 - \exp \left( - \nu \mu (T)\right),
    \label{eq:Ph}
\end{equation}
which, together with Eqs. \eqref{eq:kh4}--\eqref{eq:nu}, leads to
\begin{equation}
    k_\textrm{ko}^\textrm{obs} (T) = k_\textrm{ko}^\textrm{theor} (T) 
    \left( 1 - \frac{2}{3} c_\textrm{ad} \right)^{\frac{1}{N_\textrm{v}} t_\textrm{f} f_0 \exp \left( \frac{-E_\textrm{m}}{k_\textrm{B} T} \right)}.
    \label{eq:kh5}
\end{equation}
At temperatures above 500~K and for typical frame times $t_\textrm{f} \approx 0.5$ s, the healing probability resulting from equation \eqref{eq:Ph} is close to 1 (Fig.~\ref{fig:cross_section_full_model}), which would imply that despite their increasing creation, no vacancies can be observed.

The remaining crucial missing piece of the model is the shape of the electron probe, which leads to a statistical distribution of the positions of the created vacancies with respect to the probe position. Depending on this relationship, the time between the creation of a vacancy and its detection varies from just a few pixel dwell times to almost one frame time, with a probability distribution corresponding to the electron probe current density profile. In particular, if the "leading" tail of the electron probe, i.e. the electrons impinging on the sample where the scan has not yet reached, causes the knock-on event, we will almost immediately record the vacancy giving it little chance to heal (see Fig. \ref{fig:schematics_probe}). Conversely, if the lattice atom is ejected at a position that the beam center has already crossed by that time, the detection time will be roughly one frame time, and the vacancy very likely has already healed before it can be observed.

Replacing $t_\textrm{f}$ with the varying detection time $t_\textrm{d}$ based on a probe current density profile determined by optimizing a model of the probe in an image simulation to reproduce the observed image contrast~\cite{supplement} (Fig.~\ref{fig:schematics_probe}, approximated by a FWHM of $\sim$1.16~\AA; a simple Gaussian shape leads to qualitatively similar results) completes a final elaboration of our model that now qualitatively matches our experimental data (see Fig. \ref{fig:cross_section_full_model}). There are two unknowns in the model, namely the migration barrier $E_\textrm{m}$ and the adatom concentration $c_\textrm{ad}$, the former of which can be be estimated via first-principles simulations.

We performed nudged elastic band (NEB) calculations of the carbon adatom migration barrier using the density functional theory package \textsc{gpaw}~\cite{enkovaara_electronic_2010}. We used the finite-difference basis with grid spacing of 0.18\,\AA,  $6\times6$ graphene supercell, $6\times6\times1$ Monkhorst-Pack \textbf{k}-point mesh and convergence criterion of 0.02 eV/\r{A} for the forces. Since the inclusion of dispersion corrections has been shown to influence barrier heights~\cite{hardcastle_mobile_2013, thinius_theoretical_2014}, we used both the Tkatchenko-Scheffler (TS)~\cite{tkatchenko_accurate_2009} van der Waals (vdW) correction on top of the PBE exchange-correlation functional or an explicit treatment of vdW interactions via the C09-vdW functional~\cite{cooper_van_2010}. Our results show that inclusion of dispersion interactions has little effect on the barrier. The calculated values are 0.42 eV, 0.41 eV, and 0.39 eV for PBE, PBE-TS, and C09-vdW, respectively.

Several choices of model parameters $(E_\textrm{m}, c_\textrm{ad})$ fit our data. In Fig.~\ref{fig:cross_section_full_model}, our extended model is illustrated for migration barrier values of $0.41$ (DFT average, purple curves) and $0.33~\si{\electronvolt}$ (weighted nonlinear least-squares fit, LSQ, orange curve), with the latter better describing our experimental observations, especially at higher temperatures. The statistical uncertainty of the LSQ-fitted barrier is $0.03~\si{\electronvolt}$ for fixed values of adatom concentration and migration attempt frequency. Temperature-dependent entropic and vibrational contributions to the Gibbs free energy as well as quantum zero-point effects~\cite{henkelman_theoretical_2006} could modify the barrier and thus explain the seeming 25~\% over-estimation by DFT, but since these strongly depend on the system and diffusion path~\cite{zobelli_vacancy_2007}, we cannot estimate their relative magnitude.

Our model has one unfortunate feature: The first-order Taylor expansion of Eq.~\eqref{eq:kh5}, i.e. $(1-x)^a \approx 1 - ax$, contains the product of adatom concentration $c_\textrm{ad}$ and attempt frequency $f_0$ so that changes in their values are essentially indistinguishable. Whenever one factor is set to a seemingly reasonable value, the other will decrease to an order of magnitude that seems implausible. For $E_\textrm{m} = 0.33~\mathrm{eV}$, an adatom concentration of $10^{-3}~\mathrm{nm}^{-2}$ implies an attempt frequency of only $0.8\, \times \, 10^{8}~\mathrm{s}^{-1}$, whereas for $f_0 = 4 \times 10^{12}~\mathrm{s}^{-1}$, it leads to a very low concentration of $c_\textrm{ad} = 2.0 \times 10^{-6}~\mathrm{nm}^{-2}$. Either effective adatom concentrations are lower than we expect, or some effects missing from our model are required to explain the discrepancy. In Fig. \ref{fig:cross_section_full_model}, the values of the product $c_\textrm{ad} \times f_0$ are $(0.8 \pm 0.4) \times 10^{6}~\mathrm{nm}^{-2}\mathrm{s}^{-1}$ for the LSQ fit, and $(0.8 \pm 0.5) \times 10^{7}~\mathrm{nm}^{-2}\mathrm{s}^{-1}$ for the DFT results. For LSQ at 1073~K, the criterion $\Delta \sigma_\textrm{ko} / \sigma_\textrm{ko} = 20~\%$ was used to estimate the uncertainty of the concentration, and those of the DFT fits were set proportional to the ratios of the weighted residual variances.

We have provided the first experimental estimate of the carbon adatom migration barrier on graphene, which not only provides a useful test of widely applied modeling approaches, but also may help improve commonly used graphene growth and heat treatment techniques. Potentially, when combined with the creation and characterization of vacancies~\cite{trentino_atomic-level_2021} and the in-situ deposition of other elements, the presented approach could also be used to estimate migration barriers for other diffusing species~\cite{inani_silicon_2019}, though carbon co-diffusion will remain a complicating factor. Further experiments at higher electron energies and temperatures might give insights into additional processes such as the adatom desorption barrier and the limits of the harmonic approximation for the phonon-derived vibrational velocities.
\begin{acknowledgments}
This work has received funding from the European Research Council (ERC) under the European Union’s Horizon 2020 research and innovation programme (Grant agreement No.~756277-ATMEN) and the Vienna Doctoral School in Physics (VDS-P). Computational resources from the Vienna Scientific Cluster (VSC) are gratefully acknowledged.
\end{acknowledgments}
\bibliographystyle{apsrev4-2}
%

\end{document}